\magnification1200

\vskip 2cm
\centerline{\bf  Spacetime and large local transformations }

\vskip 1cm
\centerline{\bf   Peter West }
\vskip 1.2cm
\centerline{{ \it Mathematical Institute, University of Oxford, Woodstock Road, Oxford, OX2 6GG, UK}}
\vskip 1.2cm
\centerline{ { \it Department of Mathematics, King's College, London WC2R 2LS, UK }}
\vskip 1.7cm

\centerline { Dedicated to the memory of Lars Brink (1943-2022) for  his help and friendship over many years. }

\vskip 1.7cm
\centerline{\sl Abstract}
We argue that the existence of solitons in theories in which local symmetries are spontaneously broken requires spacetime to be enlarged by  additional coordinates that are associated with large local  transformations. In the context of gravity theories the usual coordinates of spacetime can be thought of arising in this way. E theory automatically contains such an enlarged spacetime. We propose that spacetime appears in an underlying theory when the local symmetries are spontaneously broken. 
\noindent
\vskip 6cm
email:  peter.west540@gmail.com
\vfill

\eject
\medskip
{\bf 0. Introduction}
\medskip
 A model of  spacetime was encoded in the equations of Newton but it was found not to be consistent with observations and in particular the way light travels.  It  was  replaced by the model of spacetime encoded  in the equations of Maxwell. With the advent of quantum theory and special relativity  it was realised that nature was best described by  relativistic quantum field theory which was constructed from fields whose quantisation lead to point particles.   One reason why many physicists in the 1960's did not believe in relativistic quantum field theory  was that it contained many quantities that were not measurable not least the coordinates of spacetime. 
\par
Field theories contain fields which  correspond to the  point particles whose properties they are designed to account for. However, they also have solitonic solutions and these must also be quantised. The prototype example is the magnetic monopole which appears for example in spontaneously broken SU(2) gauge theory with spin zero fields in the adjoint representation [1,2,3,4]. The motion of such objects is complicated but it is known how to find it, at least  in principle,  when the monopoles are moving slowly. The monopole solution has undetermined parameters, also called moduli, and letting these depend on time and substituting this into the action one finds their dynamics [5]. For a single monopole there are four moduli, three of these are the obvious position of the moduli in space while the fourth arises from a large gauge transformation associated with the preserved $U(1)$ symmetry. The dynamics of the monopole cannot be described by its motion as recorded by its three spatial positions, it requires in addition the four moduli. In section one we will review these developments and advocate the point of view that the underlying spacetime   really has five dimensions. 
\par
Theories of gravity also have solitons including objects extended in spacetime, that is, branes. These also have moduli which are associated with large gauge transformations. One can treat these moduli as one did for the monopole and try to find the motion of slowly moving branes. One finds in this way not only the usual coordinates that specify where the branes are  but also the world volume fields for those branes that have them, at least in the linearised approximation [6]. We will review and extend  the findings of this paper stressing in particular how even the usual coordinates of spacetime arise in this way. We will argue that the solutions of the maximal supergravity theories, and in particular the branes,  require a spacetime  that can uniquely label events and so describe the motion of solitons. Thus we require a spacetime that has coordinates 
corresponding to all the moduli that arise in the solutions. In this way one finds a spacetime that has the usual coordinates of spacetime but in addition other coordinates which arise from the other moduli. Indeed taking account the presence of all the solutions one would expect a spacetime with an infinite number of coordinates.  
\par
E theory is the non-linear realisation of the semi-direct product of the Kac-Moody algebra $E_{11}$ with its vector representation $l_1$, denoted 
$E_{11}\otimes_s l_1$. It contains all the fields of the maximal supergravity theories,  as well as an infinite number of additional fields. All these   depend on an infinite number of coordinates which arise from the vector representation.  At the lowest level are the usual coordinates of spacetime but at higher levels it includes coordinates which are those that would arise from the moduli of the solutions. Thus E theory contains the  spacetime that arises from including soliton solutions. 
\par
The non-linear realisation determines the field equations and these  contain the maximal supergravity theories  if one takes the fields to depend only on the usual coordinates of spacetime. This so far puzzling aspect is explained by fact that the supergravity theories propagate only their degrees of freedom and so do not directly  include the solitons which give rise to the additional coordinates. This step breaks the $E_{11}$ symmetry but this is to be expected as one has chosen not to take account of the brane solutions contained in the theory and also transformed by the $E_{11}$ symmetry. 
\par
The coordinates of spacetime should  uniquely label events and in particular the dynamics. As such spacetime should contain  the  coordinates which can describe the motion of the elementary particles but also those for   particles and branes  that arise as solitons. In this way one finds an extension of spacetime that is an effective spacetime in a similar sense to the way we view an effective action. We suggest that spacetime should be treated as an  observable rather than a set of parameters that appear in a theory and that it should emerge from a fundamental theory as a result of the spontaneous breaking of local symmetries. Some preliminary  thoughts in the direction of this paper were given in the paper [30]. 

\medskip
{\bf 1. The spacetime of the monopole}
\medskip
Dirac so liked the idea of a a duality symmetry between electric and magnetic fields  that he proposed the existence of objects carrying magnetic charges, that is,  monopoles.  A finite energy smooth monopole solution was found in a  SU(2) gauge theory with spin zero fields in the adjoint representation [1,2,3,4] which spontaneously break the gauge symmetry to $U(1)$. This solution has four  parameters (moduli) that can be freely varied and it is still a solution. Three of these correspond to the obvious fact that the monopole is free to sit at any point of space. The fourth moduli corresponds to a large 
$U(1)$ gauge transformation which allow the monopole to also carry electric charge. This theory also admits solutions corresponding to $N$ magnetic monopoles . These solution have $4N$ moduli which consists of the $3N$ positions of the $N$ monopoles and $N$ large gauge transformations. This is what one expect at least for widely separated monopoles. 
\par
To find the scattering of monopoles is a very complicated problem, but a way to systematically find it when the monopoles are moving at slow speeds was found by Manton [5]. The idea was to let the moduli which were constant depend on time. This did not affect the potential energy but does give the monopoles some kinetic energy. The action corresponding to this motion is found by substituting the solution with the time varying moduli into the action for the theory. 
\medskip
{\bf 1.1 The monopole solution}
\medskip
In this section it will be  instructive to review the   monopole solution emphasising  in detail some points that will be useful for the discussions in this paper. A review which has the characteristic precision and clarity is the one of Goddard and Olive [7] where the key references to this work can be found. 
The action for an SU(2) gauge theory with an adjoint scalar field is 
$$
A= \int d^4x Tr\{  -{1\over 4} F_{\mu\nu}F^{\mu\nu} -{1\over 2} D_\mu \phi D^\mu \phi-V(\phi) \}
\eqno(1.1.1)$$
where $A_\mu = A_\mu^a T^a$, $\phi =  \phi^a T^a$,  $T^a$ are the generators of SU(2) and so obey $[T^a , T^b]=\epsilon^{abc} T^c$, 
  $F_{\mu\nu}= \partial_\mu A_\nu -  \partial_\nu A_\mu + e[A_\mu , A_\nu]$ and $D_\mu\phi= \partial_\mu \phi +e [A_\mu , \phi]$. Also 
  $Tr(T_a T_b) =\delta_{ab}= (T_a T_b)$ and is the Cartan-Killing metric.  The potential is given by $ V(\phi )= {1\over 4}(Tr \phi^2 -v^2)^2$. 
The action is invariant under the gauge transformations 
$$
\delta A_\mu = D_\mu \Lambda = \partial_\mu \Lambda+ e [A_\mu , \Lambda ] , \ \ \delta \phi = e [\phi, \Lambda ]
\eqno(1.1.2)$$
\par
The electric and magnetic fields are defined by 
$E_i= F_{0i}, \ B_i= {1\over 2} \epsilon_{ijk} F_{jk}$
and in terms of these the equations of motion are 
$$
D_0 E_i+\epsilon^{ijk} D_j B_k +e[\phi , D_i \phi ]=0 ,\ \ D_i E^i+e [\phi ,D_0\phi]=0  \ \ {\rm Gauss\  Law}\ \ 
\eqno(1.1.3)$$
The Bianchi identities are 
$D_i B^i=0 , \ \ D_0 B_k -\epsilon _{kij} D^iE^j=0$. 
\par
We can write the action in terms of its kinetic energy $T$ and potential energy $V$, that is, $A= \int dt (T-V)$ where 
$$
T= \int d^3 x Tr \{ {1\over 2} E_iE^i +   {1\over 2} D_0\phi D_0\phi \} , \ \ V=  \int d^3 x Tr \{ {1\over 2} B_i B^i + {1\over 2} D_i\phi D^i\phi +V(\phi)  \} ,
\eqno(1.1.4)$$
\par
The SU(2) symmetry is spontaneously broken to U(1) when the scalar field takes a constant value such that $Tr \phi^2 =v^2$. As such the scalar fields take non-zero values at infinity and in particular on the sphere at infinity. In tangent space and in spherical coordinates the scalar contribution to the energy takes the form 
$$
-{1\over 2} \int dr d\Omega r^2  D_m\phi D^m \phi 
\eqno(1.1.5)$$ 
where $d\Omega= d\theta d\varphi \sin \theta$, $D_m=e_m{}^i D_i$ and $e_m{}^i $ is the veirbein  which  is given by $e_m{}^i = {\rm diag} ( 1,r, r\sin \theta )$. Clearly for the energy to be finite $D_m\phi$ must go to zero faster than ${1\over r}$ at infinity. This would not be the case if there was no gauge field and the fall of  $\phi$ was of the simple  form 
$$
\phi(t, r,\theta, \phi)= v+ {\phi^{(1)} (t, \theta, \phi)\over r}+\ldots 
\eqno(1.1.6)$$
Thus we can not have a finite energy solution unless $\phi$ has a fall off that is ${1\over r^2}$ or smaller. As is apparent the  fall off at infinity is most easily discussed when quantities are viewed from a  tangent space perspective and so we will below revert to tangent space when required. 
\par
However, when we have a gauge theory  the gauge covariant derivative of $\phi$ will include a term with the gauge field and demanding that $D_m\phi$  fall off faster than ${1\over r}$ we can solve for the gauge field asymptotically to find that [8]
$$
A_m = -{1\over ev^2} [\phi, \partial_m \phi]+{\phi \over v} a_m +O({1\over r^2}),
\eqno(1.1.7)$$
where $a_m$ is an arbitrary gauge field. The corresponding  field strength is given by $F_{\mu\nu}^a={ \phi^a \over v} {\cal F}_{\mu\nu}$ where 
$$
{\cal  F}_{\mu\nu}= -{1\over ev^3} Tr (\phi [\partial_\mu \phi , \partial_\nu \phi ] )+ {1\over v} (\partial_\mu a_\nu -\partial_\nu a_\mu) 
\eqno(1.1.8)$$
\par
We observe that the field strength is in the direction of $\phi^a$ which is asymptotically the direction of the symmetry breaking by the scalar fields and so ${\cal F}_{\mu\nu}= {\phi^a \over v} F_{\mu\nu}^a$. For the one monopole solution $\phi^a$ points in the $r^a$ direction. 
\par
One can check that $\partial_\mu {\cal  F}^{\mu\nu}=0=\partial_{[\mu} {\cal  F}_{\nu\rho]} $  and so at infinity we have in effect a Maxwell field. The associated electric and magnetic fields are given by 
$$
{\cal E}_i= Tr (\phi {\cal  F}_{0i}), \ {\rm and } \ {\cal B}_i = {1\over 2}\epsilon _{ijk} Tr (\phi {\cal  F}^{jk})
\eqno(1.1.9)$$
The corresponding magnetic and electric charges are  given by 
$$
g= \int _{S^2} {\cal B}_i dS_i= -{1\over 2 e v^3} \int _{S^2}\epsilon^{ijk}Tr (\phi [\partial_j \phi , \partial_k \phi ] )dS_i
\equiv {4\pi\over e} N ,\ \ q=  \int _{S^2} {\cal  E}_i dS_i
\eqno(1.1.10)$$
 Here $N$ is a topological number which can only take integer values. As the symmetry is spontaneously broken the  scalar field must have a non-zero value  on the sphere at infinity. However, the scalars are also valued in SU(2),  and so on a sphere,  and as such one has the mapping of this  sphere onto the sphere at infinity. The number of times the scalar wraps wound the sphere at infinity is the number $N$. 
\par
In tangent space and  in spherical coordinates the above expression for the magnetic charge takes the form 
$$
 g= -{1\over 2 e v^3} \int _{S^2} d\Omega r^2 \epsilon^{r\theta \varphi}Tr (\phi [\nabla_\theta \phi , \nabla_\varphi \phi ] )
= -{1\over 2 e v^3} \int _{S^2}d\Omega Tr (v[\nabla_\theta \phi^{(1)} , \nabla_\varphi \phi ^{(1)} ] )
\eqno(1.1.11)$$
 where $\nabla_m= e_m{}^i \partial_i $ is the ordinary derivative in tangent space, that is, without the gauge field. Looking at the first equality it is clear that to have a non-zero finite magnetic  charge $g$, or equivalently $N$, $\phi$ must fall off as in equation (1.1.6) and as we explained above this is only possible for a finite energy solution if we have a gauge field present. 
 The gauge field falls off as in equation   (1.1.7).  In the last equality in equation (1.1.11) we have used the fall off of $\phi$ of equation (1.1.6). 
 \par
The energy of a solution is given by 
$$
E= T+V=  \int d^3 x Tr \{ {1\over 2} E_iE^i +   {1\over 2} D_0\phi D_0\phi + {1\over 2} B_i B^i + {1\over 2} D_i\phi D^i\phi +V(\phi)  \}
$$
$$
\geq   \int d^3 x Tr \{ {1\over 2} E_iE^i +{1\over 2} B_i B^i + {1\over 2} D_i\phi D^i\phi   \}
$$
$$
=
{1\over 2} \int d^3 x  Tr\{(E_i- D_i \phi \sin \theta) ^2+(B_i- D_i \phi \cos  \theta)^2\}=  v(q\sin \theta + g \cos \theta) \geq v \sqrt {g^2+ q^2}
\eqno(1.1.12)$$
In getting to the second to last line we have  used the Bianchi identity $D_iB^i=0$ and the equation $D_i E^i$ which follows if $D_0\phi=0$ as well as equation (1.1.10). 
Equality follows in equation (1.1.12) if we take 
$$
D_0\phi=0, \  E_i=0,\  V(\phi)=0,\ B_i=D_i\phi
\eqno(1.1.13)$$
\par
These conditions are   satisfied by the  monopole solution and in this case it has the mass 
$$
M= {1\over 2} \int d^3 x  Tr \{B_i B^i +D_i\phi D^i\phi\} = \int d^3 x  Tr \{B_i B^i \}= v\sqrt {g^2 + q^2} 
\eqno(1.1.14)$$
The mass of one monopole is given by $M_1=  {4\pi v\over e}$ where we have used the fact that the monopole has magnetic charge $g={4\pi\over e}$ from equation (1.1.10). We have necessarily given a much shortened account and we refer the reader to the reviews of reference [7] for more details and the original references. 
\medskip
{\bf 1.2 The monopole moduli and its motion }
\medskip
In this section we will describe the moduli that monopole solutions possess and explain how these can be used to derive the low energy motion of monopoles [5]. For a review of this material which covers many parts of our discussion see references [9] and [10].  The above solution of the one monopole has four undetermined parameters. Three of these are obvious,  they are just the position of the monopole in space. This must be the case as the theory is invariant under translation symmetry $x^i\to x^i+a^i$ with $a^i$ a constant. However the fourth is rather subtile and is associated with gauge transformations. In general gauge transformations do not lead to physical effects however, there can be gauge transformations that die off sufficiently slowly at infinity that they lead to non-trivial effects at infinity and in particular when they occur in quantities that are  integrated over the sphere at infinity. These are know as large gauge transformations. 
\par
Let us consider a change in the solution by a gauge transformation of equation (1.1.2) but with the gauge parameter $\Lambda = {\chi \phi \over v}$ where $\chi$ is a constant. The solution then   takes the form 
$$
A_i \to  A_i + \delta A_i = A_i +D_i ({\chi \phi\over v}), \ \  \phi\to  \phi + \delta \phi =
\phi +e [\phi , {\chi \phi  \over v}]= \phi  ,\ \  A_0\to A_0 +\delta A_0 = A_0
\eqno(1.2.1)$$
The last equation follows from the fact that the solution has $D_0\phi=\partial_0\phi +[A_0 , \phi]=0 = \partial_0\phi$ from equation (1.1.13) and due to the fact that we are working  in the gauge $A_0=0$. We note that the gauge parameter goes at infinity as 
$\Lambda (t,r, \theta, \varphi ) ={ \chi \over v} (v+ {\phi ^{(1)}(t, \theta, \varphi )\over  r} +O({1\over r^2}))$ and so has a the same non-trivial fall off as $\phi$.  
For the change of equation (1.2.1) we find that $E_i=0$, the Gauss law of equation (1.1.3) holds and so does  the last condition of equation (1.1.13) as must be the case as it is just a gauge transformation. 
\par
Thus the solution has four arbitrary parameters,  or moduli, $a^i$ and $\chi$. We will now let these  moduli depend on time, more precisely we let   $\chi \to \chi (t)$. Under this   the solution changes by 
$$
A_i \to A_i + \delta A_i =   A_i +D_i ({\chi (t)\phi \over v})  , \ \  \phi\to  \phi + \delta \phi = 
\phi   ,\ \  A_0\to  A_0 +\delta A_0 = A_0 + D_0 ({\chi (t) \phi \over v}) - {\dot \chi \phi \over v}=A_0
\eqno(1.2.2)$$
However, we also   enforce the gauge condition $A_0=0$ and so the change is no longer a gauge transformation. With these changes we now have an electric field 
$$
E_i=\dot \chi {D_i \phi\over v} = {\dot \chi B_i \over v}
\eqno(1.2.3)$$ 
The other conditions of equation (1.1.13) still  hold as $D_0\phi=0$ and  
$$
\delta B_i ={1\over 2} \epsilon^{ijk} \delta F_{jk}= {1\over 2} \epsilon^{ijk} [ [D_j , D_k ] ,{ \chi \phi \over v}] = e [B_i , {\chi \phi\over v}] 
= e[D_i \phi, {\chi \phi\over v} ] = \delta (D_i  \phi )
\eqno(1.2.4)$$
We can check that  the Gauss law still also holds as
$$
D_i E^i+e [\phi , D_0\phi ]={\dot \chi D_iD^i\phi \over v}=0 
\eqno(1.2.5)$$ 
as $D_iB^i= 0=D_i D^i \phi$. 
\par
 Thus the effect of the change of equation (1.2.2) is to introduce an electric field  and so give the monopole an electric charge. This is a contribution to the kinetic energy of the monopole, but  it does not increase the potential energy. To see this we examine  equation (1..12) and notice  that the potential energy vanishes if   $B_i=D_i\phi$ apart from a topological term which is unchanged. 
\par
The motion of  a slowly moving monopole is found by  substituting the changed solution of equation (1.2.2)  into the action of equation (1.1.1). Carrying out the integrations over space, we will find an action that is an integral over time and depends on the moduli $a^i(t)$ and $\chi(t)$ and their time derivatives. The dependence on $\chi$ arises from the electric field and the contribution to  the action is 
$$
{1\over 2 } \int dt d^3 x Tr \{ E_i E^i\} = {1\over 2 v^2}\int dt\dot \chi^2 \int d^3 x Tr \{B_i B^i\}=  {1\over 2 v^2}M_1 \int dt\dot \chi^2
\eqno(1.2.6)$$ 
\par
We will now derive the  contribution to the action due to  the moduli $a^i$ in  detail as,   although this is well known,  this author did not find an explicit derivation in the literature. We undertake a coordinate change $x^{i\prime}= x^i + a^i (t)$, $x^{0\prime }= x^0$ under which 
$$
\delta A_0= \dot a^k A_k , \ \ \delta A_i = a^k \partial _k A_i
\eqno(1.2.7)$$
However this does not preserve our gauge choice, $A_0=0$, and so we carry out a compensating gauge transformation with parameter $\Lambda = -\dot a^k A_k$ whereupon we have the changes 
$$
\delta A_0= 0  , \ \ \delta A_i = a^k F_{k i} \ \ {\rm and } \ \ \delta \phi= a^k D_k \phi
\eqno(1.2.8)$$
In the last equation we have carried out these same steps for the scalar fields $\phi$. Under these changes $\delta E_i = \epsilon_{ijk}  \dot a^j B^k$   and $\delta ( D_0\phi )= \dot a^k D_k \phi$ and substituting these into the Gauss law of equation (1.1.3) we find that it still holds. 
\par
Under the changes of equation (1.2.8) we find that $\delta B_i = a^k D_k B_i$ and $ \delta (D_i \phi )= a^k D_k D_i \phi$  and as a result the condition $B_i = D_i \phi$ still holds. Following the same argument as above we conclude that these  changes do not alter the potential energy.  
\par
However the variations  of equation (1.2.8)   do change    the kinetic energy as follows  
$$
{1\over 2} \int dt d^3x  \{  \dot a^i \dot a_i  B^j B_j - \dot a ^i B_i \dot a^j B_j + \dot a^i D_i \phi   \dot a^j D_ j \phi\}
={1\over 2}  \int dt  \dot a^i \dot a_i  \int d^3 x B^j B_j = {M_1\over 2}   \int dt  \dot a^i \dot a_i 
\eqno(1.2.9)$$
\par
Thus letting the moduli depend on time and substituting this into the action of equation (1.1.1) we find that it becomes  
$$
{M_1\over 2} \int dt( \dot a_i \dot a^i + {1\over v^2} \dot \chi^2 )
\eqno(1.2.10)$$ 
Thus the motion is given by $\ddot \chi =0=  \ddot a^i$ and so the monopole moves with constant velocity in space and also in $\chi$. We will not write $a^i$ and $x^i$ to make its meaning obvious. 
\par
The momenta corresponding to this action are given by 
$$
p^i= M_1 \dot x^i =-i\hbar{\partial\over \partial x^i}, \ \ p=  {M_1\over v^2} \dot \chi = {4\pi \over ev} \dot \chi =-i\hbar{\partial\over \partial \chi}
\eqno(1.2.11)$$
where we have also implemented the quantisation conditions. 
The Hamiltonian is given by 
$$
H= {p_ip^i \over 2M_1}+ {ev\over 8\pi} p^2 =- {\hbar^2 \over 2M_1} {\partial ^2 \over \partial x^i{}^2} -  {ev\hbar^2 \over 8\pi}{\partial^2\over \partial \chi^2}
\eqno(1.2.12)$$
\par
The electric field of the monopole at infinity is given by ${\cal E}_i= Tr({E_i \phi )\over v}= {\dot \chi  \over v}{\cal B}_i $ and the 
 corresponding electric charge as defined by  equation (1.1.10)  is given by 
   $$
   q= {\dot \chi g \over v} = {4\pi \over e v}\dot \chi = p= -i \hbar {\partial\over \partial \chi}
   \eqno(1.2.13)$$
   where in the last step we have used the quantum expression for the momentum. 
The electric charge must obey the Dirac quantisation condition $gq={4\pi \over e}p= 4\pi n$ and so we conclude that the momenta $p$ must take the values $p=ne$ where $n$ is an integer. As a result the monopole has the original magnetic charge $g={4\pi\over e}$ and also an electric charge $ne$. We note that the equation of motion of $\chi$ ensures that the monopole has constant electric charge. 
\par
The  wavefunction $\psi$ of the quantum monopole  depends on the the coordinates $t, x^i , \chi$, that is, $\psi (t,  x^i , \chi )$ and will obey the Schr\" odinger equation $H\psi = i\hbar{\partial \psi \over \partial t}$. If we take a stationary wavefunction, that is,  $\psi= e^{-{iEt\over \hbar}} \hat \psi$ then we have 
$H\psi= E\psi$ and if we then take a plane wave configuration, that is,  $\hat \psi = e^{{ip_ix^i\over \hbar}} e^{{ip\chi\over \hbar}}$ then we have the energies 
$$
E= {p_ip^i\over 2 M_1} +  {e^3 n^2 v\over 8\pi}
\eqno(1.2.14)$$
Adding the rest mass $M_1$ and taking the momentum $p_i=0$ we find that it agrees with the mass derived in equation (1.1.14), that is, 
$M= v\sqrt {g^2 + q^2}= vg +{vq^2\over 2g}= {4v\pi\over e}+  {e^3 n^2 v\over 8\pi}+\ldots $. 
\par
The generator of electric charge of equation (1.2.13) obviously generates  a shift in the $\chi $ coordinate $\chi\to \chi +w$ where $w$ is the parameter of the transformation. This in turn results in a gauge transformation of $A_i$ of the form $A_i\to A_i+ {w D_i \phi \over v}$, that is, a gauge transformation with parameter $\Lambda= {w \phi \over v}$. Thus the combination $A_i + {\chi D_i \phi \over v}$ is invariant under this transformation. 
\par
The solution with $N$ monopoles has $4N$ moduli, denoted $z_\alpha$. This is to be  expected when we think about the monopoles being widely separated. Of these $3N$ are the coordinates of the positions of the moduli and $N$ are large gauge transformations.  In this case the slow motion of the $N$ monopoles is found by letting the moduli $z_\alpha$ depend on time $z_\alpha \to z_\alpha (t) $ and evaluating the action for the solution with this substitution, that is, for $A_i (x^j , z_\alpha (t))$ and $\phi  (x^j , z_\alpha (t))$. One can show that the  potential energy does not depend on the  moduli and so we need only consider the kinetic terms are given in equation (1.1.4). In the gauge $A_0$ we then  find the action for the motion to be 
$$
{1\over 2} \int dt d^3 x \{ \dot A_i  \dot A_i +\dot \phi \dot \phi \}= {1\over 2} \int dt \dot z_\alpha g_{\alpha \beta} \dot z _\beta
\eqno(1.2.15)$$
where 
$$
g_{\alpha \beta} = \int d^3 x \{ \partial _\alpha A_i  \partial _\beta A^i +  \partial _\alpha \phi  \partial _\beta \phi  \}
\eqno(1.2.16)$$ 
In doing this one has  to add a compensating gauge transformation as we did above. The classical motion of the monopoles has been extensively studied see for example [11]. 
\par
One can also quantise this  action in a way that is similar to that for one monopole but this is much more complicated. This approach has been used to calculate the quantum scattering of two monopoles [10] and it played a key role in our  understanding of monopoles in the maximally supersymmetric Yang-Mills theory and  its electromagnetic duality symmetry,   see reference [12].  

\medskip
{\bf 1.3 The consequence of monopole motion for spacetime.  }
\medskip
The role of spacetime is to label events. In practice we need to assign coordinates to events in such a way as to label them uniquely. The dynamics of objects is then given by the time evolution of their coordinates. As illustrated by the action of equation (1.2.10) for the slow motion of one monopole it is clear that their motion can not be described in term of the three coordinates of space $x^i$ but one requires the additional coordinate $\chi$. For $N$ monopoles the motion is geodesic motion on moduli space which has dimensions $4N$ whose coordinates are the $3N$ spatial positions of the $N$ monopoles and the $N$ $\chi$ coordinates [11]. Although for very widely spaced monopoles the $\chi$ coordinates do not play much role they are essential when the monopoles are close to each other or when the come close when scattering. 
\par
It is well known that to find the slow motion of the monopoles one needs to take account of the $\chi$ moduli. One might regard them  as something that is associated with the internal behaviour of monopoles, however, in this paper we will take the view point that the monopoles are probing spacetime in a way which the purely electrically charged particles are not and in the process they are finding that spacetime has one more coordinate than we usually use, namely the $\chi$. Thus we regard spacetime as being enlarged and having five dimensions. In all we then have   time, the three spatial coordinates and the $\chi $ coordinate. That the moduli space of $N$ monopoles has $4N$ dimensions is consistent with this interpretation.  
\par
This way of viewing the situation  is particularly natural in the context of the maximally supersymmetric Yang-Mills theory as we now explain. Let us take the theory to have an SU(2) gauge group and then it has the from of the theory we considered above except that there are more scalars and also fermions. It therefore has the above monopole solutions and we can apply the above analysis to find monopoles with electric $q$ as well as magnetic charges $g$. Of course the theory also possess the supersymmetry algebra and the anti commutator of the supersymmetry charges takes the form. 
$$
 \{ Q^i_\alpha . Q^j_\beta \}= 2 (\gamma^a C)_{\alpha \beta} P_a \delta^{ij} +\epsilon ^{ij} v^2((q +\gamma_5 g)C)_{\alpha \beta}
\eqno(1.3.1)$$
See reference [12] for some reviews. The algebra possess the central charges which are in fact the electric and magnetic charges as shown above. Indeed the BPS condition is 
$p^2+v^2(g^2 +q^2) =0$ which we recognise as being the same as equation (1.1.12)
\par
The $N=4 $ supersymmetry algebra consists of the generators $P_\mu, q, g , Q_{\alpha i}$ as well as  the Lorentz generators $J_{\mu\nu} $ and the SO(6) generators of the internal symmetry. The spacetime  in which this theory lives is the coset space of the supersymmetry algebra divided by the Lorentz group and the internal symmetry group. The cosets correspond to the group element  
$$
e^{x^\mu P_\mu + \chi q + \lambda  g} e^{\theta^{\alpha i} Q_{\alpha i}}
\eqno(1.3.2)$$
which leads to a spacetime with the coordinates $x^\mu ,  \chi ,  \lambda , \theta^{\alpha i}$. The corresponding superfields will contain the anti-commuting $ \theta^{\alpha i}$ which can be eliminated in a rather trivial way from the action by integration however, one is left with fields that depend on $x^\mu ,  \chi $ and  $ \lambda $. From this perspective the  $\chi$ coordinate occurs in essentialy the same way as the usual coordinates $x^\mu$ of spacetime and it is very natural to  regard $\chi$ as a coordinate of spacetime just like $x^\mu$. We note the appearance of another coordinate $\lambda$ associated with the magnetic charge. 
\par
 Carry out a transformation $g_0=e^{wq}$ on the group element of equation (1.3.2)   would shift the $\chi$ coordinate $\chi \to \chi +w$. This is indeed the shift in the $\chi$ coordinate under the generator of electric charge $q$ that we introduced in section 1.2. As a result we can identify the coordinate $\chi$ that occurs in the group element of equation (1.3.2) with the $\chi$ coordinate that arises as a moduli in section 1.2. 
\par
The maximally supersymmetry Yang-Mills theory has the electric and magnetic duality symmetry proposed by Montenon and Olive [13] for a review see the book [12]. From the view point of the dual theory the monopoles appear as fundamental particles and the purely electrically particles as solitons. These solitons can then carry magnetic  charges as well as electric charges which arise from large gauge transformations associated with the magnetic charges. Just as the $\chi $ coordinate arose in the original formulation of the theory these large gauge transformations will introduce an additional coordinate  $\lambda $. This must be the additional  coordinate that also appears in the group element of equation (1.3.2). 
The duality symmetry will rotate the electric and magnetic charges into each other and as a result transform the $\chi$ and $\lambda$ coordinates into each other.  
\par
Regarding the $\chi$ and $\lambda$ coordinates as part of spacetime is very more natural if we recall the papers  [14] and [15]. In these papers it was argued that the many features of the monopoles that occur in the $N=4$ supersymmetric Yang-Mills theory can be seen  as arising from the uplift of this theory to six dimensions. In particular the  central charges in  equation (1.3.2) arise  as the components of the momenta in the fifth and sixth directions [14, 15]. In this way the electric and magnetic charges have a six dimensional origin [14] including their  topological nature  [15] . 
\par 
If we were to second quantise the  action describing the monopole motion resulting from the Manton approximation, discussed above, we would find a relativistic quantum field theory whose fields $\psi$ would depend on the usual coordinates of spacetime but also on the coordinate $\chi$. One might expect that it would obey, at the linearised level, a condition of the very generic  form  
$$
(-\partial_a \partial^a + v^2 e^2{\partial^2\over \partial\chi^2}  + ({4\pi v\over e})^2 )\psi=0
\eqno(1.3.3)$$
This equation should describe not only the monopoles but also the dyons for a suitable set of fields $\psi$. Indeed just such an equation generalised for two monopoles and the corresponding moduli space has been used to account for the quantum behaviour of two such monopoles [10]. Of course one could include the $\lambda$ coordinate in a similar way to the $\chi$ coordinate.
\par
When we are considering only the physics of the original electrically charge particles we have no need to introduce the $\chi$ and $\lambda$ coordinates and one can just use the usual coordinates of spacetime. However, if we want to account for the behaviour of the monopoles  we need the $\chi$ coordinate and if we want a duality symmetric formulation we will also need the $\lambda$ coordinate. If there exists a formulation which includes the elementary and monopole particles on an equal footing and encodes duality symmetry is unknown.


\medskip
{\bf 2. The spacetime of branes  }
\medskip
In this section we will examine solutions in supergravity theories and process them in much the same way as we did for the monopole.  In particular we will identify the undetermined parameters, the moduli,  of   the solutions and by promoting these to be spacetime dependent we will find the low energy motion of the solitons. The difference with the monopole is that the solutions are branes and so they have a world volume that is extended in space as well as time. The moduli will arise as large gauge transformations and we let them depend on  the coordinates of the spacetime  swept out by the brane, that is, the world volume coordinates. This should not change the potential energy of the branes but it will change the kinetic energy.  Making this change  in the solution and substituting it into the action we will  find an action that depends on the moduli and determines the low energy motion of the branes. For the branes in M theory one  should take the actions of the maximally supergravity theories in ten and eleven dimension and make the substitution. The strategy underlying such an approach was given in reference [6],  albeit in the language of zero modes and Goldstone particles, and worked out   in some detail for   the lowest order contributions to the equations of motion for the M5 and D3 branes. We will review this work, generalise it other branes and give a more detailed treatment using the action to derive the dynamics. 
\par
Such an approach was also used to derive the low energy effective action of $N=2$ Yang-Mills theory. It was realised that when the M5 brane is wrapped on a two dimensional Riemann surface it could be used to compute such effects [16]. At the  intersections of the  five brane we find three branes living within the five brane and taking the moduli that appear in  this three brane  solution to be spacetime dependent  we find coordinates that  describe the motion of the slowly moving three branes. In this way one recovers  in three pages [17] the Seiberg-Witten effective action [18]. 
\par
Let us first recall the properties of the half BPS solutions of the maximally supergravity theories in ten and eleven dimensions. We refer  the reader to the review in chapter thirteen  of the book of reference [19] and we  use almost  the same conventions.  We will consider a brane solution in the  theory  in $D$ dimensions with spacetime coordinates $x^{\mu}= (x^j, y^n)$ which has the form 
$$
A={1\over 2\kappa_D^2} \int d^D x\sqrt{-g} (R-{1\over 2}\partial^\mu\phi\partial_\mu\phi 
-\sum_n {1\over 2 (n+2) ! } e^{a_{n+2}\phi} F_{\mu_1\ldots \mu_{n+2 }}  F^{\mu_1\ldots \mu_{n+2 }} )  
\eqno(2.1)$$
where $F_{\mu_1\ldots \mu_{n+2 }}=(n+2) \partial_{[\mu_1}A_{\ldots
\mu_{n+2 }]}$ and $R_{\mu\nu}{}^\mu{}_\rho=R_{\nu\rho}$,
$R=g^{\mu\nu}R_{\mu\nu}$. 
The constant $\kappa_D$ is related to the usual gravitational constant by $2\kappa_D^2=16\pi G_{D}$. The sum over $n$ is not over all $n$ and it is  different in the different theories. The precise relation to the maximal supergravity theories in eleven and ten dimensions is given in chapter thirteen of reference [19]. The Chern-Simmons term is not shown as it vanishes for the solutions we consdier. 
\par
The $p$ brane solution has the metric 
$$
ds^2= dx^\mu g_{\mu\nu} dx^\nu = N^{2\alpha} dx^i \eta_{ij} dx^j + N^{2\beta} {dy^n }\delta_{nm} dy^m
\eqno(2.2)$$
where $N$ is a function of $y^2=y^n y_n$ and $\alpha= -{(d-2)\over \Delta}$ and $\beta ={(p+1)\over \Delta}$ . The space transverse to the brane has dimension $d= D-p-1$,  $N= 1+ {1\over (d-2) }\sqrt {{\Delta \over 2(D-2)}}{Q\over  r^{d-2}}$ with $Q$ being a constant and 
$\Delta= (p+1) (d-2) + a^2_{p+2} {(D-2)\over 2}$. The field strength is zero except for the components 
$$
F_{i_1\ldots i_{p+1} n}=\pm \sqrt {{2(D-2)\over \Delta}} \partial_n N^{-1} \epsilon_{i_1\ldots i_{p+1} }
\eqno(2.3)$$
In eleven dimension $\Delta=18$ while in ten dimensions $\Delta=16$. The dilaton in the  ten dimensional theories  has the value $e^\phi= N^{a_{p+2}(D-2)\over \Delta}$. In both of the ten dimensional maximal supergravity theories $a_{p+2}= \eta{3-p\over 2}$ where $\eta=1$ and $-1 $ for fields in the R-R and NS-NS sectors respectively. The constants $\alpha$ and $\beta$ obey the relations  $(p+1) \alpha + (D-p-3)\beta=0$ and $(p-1) \alpha + (D-p-1)\beta=1$ which imply that $2(\alpha-\beta)=-1$. 
The theory of equation (2.1) has the usual gauge and  diffeomorphisms, or local translations,  symmetries which are given by 
$$
\delta A_{\mu_1\mu_2\ldots \mu_{p+1}}= (p+1) \partial_{[\mu_1}\Lambda_{\mu_2 \ldots \mu_{p+1}]} ,\ \ {\rm and }  \ \ 
\delta h_{\mu\nu}= \xi^\lambda \partial_\lambda h_{\mu\nu}+ \partial_\mu \xi^\lambda h_{\lambda \nu}++ \partial_\nu \xi^\lambda h_{\mu\lambda }
\eqno(2.4)$$
where we have written the metric as $g_{\mu\nu} = \eta_{\mu\nu} + h_{\mu\nu}$.  
\par
As for the monopole we will take rather specific gauge transformations  and local translations  which are large in the sense that they die off at infinity at a sufficiently slow rate so as to give effects which are physically measurable. Within the parameters of these transformations we will identify the moduli which we will then take to depend on the world volume coordinates of the brane. Such a path has already been  followed by the authors of reference [6]. This paper gave the general strategy and computed in some detail from the equations of motion the linearised equations that the moduli satisfied for the M5 brane and the D3 brane. In this way they were able to show that the moduli could be identified with the variables  that are used to derive the dynamics of these branes including their world volume fields. We will follow this work but be more general, detailed and take an action based approach. We will also elaborate  the points suited to the purposes of this paper. 
\par
We begin by considering the gauge transformations of equation (2.4). Following [6] we consider a  gauge transformation with the parameter 
$$
\Lambda_{i_1\ldots i_{q}}= N^{-1} B_{i_1\ldots i_{q}}
\eqno(2.5)$$
with all other components being zero. Here $B_{i_1\ldots i_{q}}$ is a constant. We note that there are several  gauge fields in the action of equation (2.1) apart from the $p$ form field that supports the $p$ brane and the above gauge transformations may arise from any of  these. As such $q$ may not be equal to $p$. Given the gauge parameter of equation (2.5), the only transformation of the gauge field that is non-zero is 
$\delta A_{n i_1\mu_2\ldots i_{q}}= \partial_n N^{-1}  B_{i_1\ldots i_{q}}$
We can identify the  $B_{i_1\ldots i_{q}}$  as moduli as it occurs in a gauge transformation that must take a solution into the same solution provided we deal with gauge invariant quantities,  such as the field strength. 
\par
Now we take the moduli, that is, the $B_{i_1\ldots i_{q}}$ to depend on the world volume coordinates of the brane, namely  $B_{i_1\ldots i_{q}} \to B_{i_1\ldots i_{q}}(x^i)$. We then take the gauge transformations to be given by 
$$
\delta A_{n i_1\ldots i_{q}}= \partial_n N^{-1}  B_{i_1\ldots i_{q}} , \ \ \delta A_{ i_1\ldots i_{q+1}}=0, \ \ \delta A_{n m i_1\ldots i_{q-1}}=0, \ldots 
\eqno(2.6)$$
It is important to note that if we were to simply take the $B_{i_1\ldots i_{q}}$ that appear in the gauge parameter of equation (2.5)   to depend on the world volume coordinates   then, according to equation (2.4),     $\delta A_{ i_1\ldots i_{q+1}}$ would be non-zero. However,   we will  take it to be zero and so in this step we are no longer carrying out a gauge transformation. In the case of the monopole this is equivalent to taking $\delta A_0=0$,  as we did in equation (1.2.1), although in this case it is also zero as we maintain the gauge choice. In the situation we have here we do not need to make a gauge choice as we can work, in effect,  with the field strength and gauge fields modulo their gauge transformations. 
The only non-zero field strength is 
$$
F_{ n i_1\ldots i_{q+1}}= \partial_n N^{-1} G_{ i_1\ldots i_{q+1}}
\eqno(2.7)$$
where $G_{ i_1\ldots i_{q+1}}= (q+1) \partial_{[ i_1}B_{\ldots i_{q+1]}}$.
\par
We now substitute the solution with the world volume dependent moduli into the action of equation (2.1) to find the result 
$$
e\int d^{p+1} x (-{1\over 2 (q+1)!} G_ {i_1 \ldots i_{q+1}} G_ {j_1 \ldots j_{q+1}}\eta^{i_1j_1}\ldots \eta^{i_{q+1} j_{q+1} } )
\eqno(2.8)$$
where 
$$
e= {1\over 2\kappa_D} \int d^d y \partial _n N \partial_m N \delta ^{nm} N^{-2\alpha(q+1) -4}N^{{a_{p+2}a_{q+2}\over 2}}
\eqno(2.9)$$
and we have used that $\sqrt {-\det g}= N^{2\beta}$. Thus we find a $q+1$ world volume gauge field. 
\par
Let us now apply the above analysis to the eleven dimensional supergravity theory which has a three gauge field $A_{\mu_1\mu_2\mu_2}$. We take the gauge symmetry to have the parameter $\Lambda_{i_1i_2}= N^{-1} B_{i_1i_2}$ and so we will find a two form world volume gauge field. For the M2 brane this would be a space-filling brane which leads to  no degrees of freedom.  However, for the M5 brane we find the two form gauge field that is know to be part of its  dynamics and was recovered from this viewpoint in reference [6]. 
\par
For the IIA theory we have gauge fields corresponding to branes with  $p=0,1,2$ and there are also their  duals $p=4,5,6$ as well  an 8 brane. However,  only the gauge fields corresponding to   $p=0,1,2$ appear in the action of equation (2.1), that is, the  gauge fields $A_\mu$, $A_{\mu_1\mu_2}$ and  $A_{\mu_1\mu_2\mu_2}$.  For the gauge symmetry of the  field $A_{\mu_1\mu_2}$  we can take $\Lambda_{i}= N^{-1} B_i$. If we do this  we find a one form world volume field for all of the branes and in particular we  would find the field content of the  D branes for $p=0,2,4,6,8$. If we take $\Lambda_{i}= N^{-1} B_{i_1 i_2}$ then we have a two form world volume field which is the one which appears in the NS-NS five brane. Which world volume field appears in which brane can be found by doing a count of the degrees of freedom so as to ensure that the number of bosonic and fermonic degrees of freedom are equal on shell. Of course it could be that there are additional non-supersymmetric branes that appear in this way and it might be interesting to study this possibility. 
\par
The Chern-Simmons term is not included in the action of equation (2.1) as it vanishes for the solutions we have been discussing. However it might play a role when the moduli are activated. In eleven dimensions it  is given by 
$$
{1\over 2\kappa^2} \int d^{11}x {1\over (12)^4} \epsilon ^{\mu_1\ldots \mu_{11}} F_{\mu_1\ldots \mu_4} F_{\mu_5\ldots \mu_8} A_{\mu_9\mu_{10} \mu_{11}}
\eqno(2.10)$$
If we are considering a brane that is supported by one of the fields in the action of equation (2.1) then the corresponding non-vanishing field strength of the $p$ brane of equation (2.3) and the moduli containing field strength of equation (2.7) have predominantly world volume  indices. As such   the Chern-Simmons term vanishes as we have repeated indices on the epsilon symbol. However, if we have a $p$ brane that is dual to these it can be non-zero. Such is the case for the five brane in eleven dimensions. In  this case it is easier to consider the equation of motion which is of the form 
$$
\partial_\nu (\sqrt {-\det g} F^{\nu\mu_1\mu_2\mu_3} ) + {1\over 12.12} \epsilon^{\mu_1\mu_2\mu_3\nu_1\ldots \nu_8}
F_{\nu_1\ldots \nu_4}F_{\nu_5\ldots \nu_8}=0
\eqno(2.11)$$
For the five brane $F_{n_1\ldots n_4}= \epsilon _{n_1\ldots n_4 b}\partial_b N^{-1}$,  using the expression of equation (2.6) with $q=2$ and taking $\mu_1=i_1, \mu_2=i_2$ and $\mu_3=i_3$ we find that 
$$
G_{i_1i_2i_3}+ {1\over 6} \epsilon _{i_1i_2i_3}{}^{j_1j_2j_3}G_{j_1j_2j_3}=0
\eqno(2.12)$$ 
In other words we recover the well known fact that the world volume two form is self-dual. 
\par
Any solution occupies a particular position in spacetime and so breaks translation symmetry. The monopole  spontaneously breaks  the rigid translation symmetry of the gauge  theory but a soliton in a  theory of gravity  spontaneously breaks  local translations. As a p brane solution has an extent in space it does not break all of the local translations,  but only those in the directions that are  transverse to the brane. As such  we take  local translation  to be of the form [6]
$$
\xi^n= N^{-1} z^n , \ \ \xi^i=0 
\eqno(2.13)$$
where $z^n$ are constants. Under this transformation 
$$
\delta h_{ij}= 2\alpha  \eta_{ij} N^{2\alpha-2 }z^n\partial_n N , \ \delta h_{in}= 0 , \ 
$$
$$
\delta h_{nm}=  N^{2\beta-2} (2\delta_{nm}\beta z^r\partial_r N -   z_n \partial_m N - z_m \partial_n N )
\eqno(2.14)$$
 It is a symmetry of the theory and so we can identify the $z^n$ as moduli. 
 \par
 We now let the moduli depend on the world volume coordinates of the p brane, that is  $z^n \to  z^n(x^i)$,  but we take the transformations of the metric to be given by equation (2.14). The transformation of $h_{in}$ under a local translation  (2.13) with $z^n (x^i)$ is  non-zero but we take it to be zero and as a result we are no longer carrying out  a local translation. 
 \par
 When we substitute the solution into the action of equation (2.1) we do not need to keep terms with no derivatives acting on $z^n$ as these must vanish as they correspond to a local translation under which the action is invariant. We will  only compute the lowest order term which is bilinear in  $z^n$  and so  has two world volume derivatives. The variation of the gravity term in the action is of the form 
 $$
 {1\over 2\kappa _D} \int d^D x \{ \delta (\sqrt {-det g}) g^{\mu\nu} \delta R_{\mu\nu}+  \sqrt {-det g}\delta g^{\mu\nu} \delta R_{\mu\nu} + 
  \delta (\sqrt {-det g}) \delta g^{\mu\nu} R_{\mu\nu} \}
  \eqno(2.15)$$
 Using equation (2.14) we find that 
 $$
  \delta (\sqrt {-det g}) = N^{2\beta -2} (\delta +2\alpha) z^m\partial_m N
    \eqno(2.16)$$
where $\delta = \alpha (p-1) +\beta(D-p-1) -1$ which  vanishes for the branes we are considering. 
Also we find  that [6]
$$
\delta R_{ij}= - \alpha   \eta_{ij} N^{-2} \partial_k \partial^k  z^r \partial_r N - N^{-2}\partial_i\partial_j z^r\partial_r N  \delta ,\ \ 
$$
$$
\delta R_{mn}= - N^{-2(\alpha-\beta) -2} \beta \delta _{mn} \partial_k \partial^k z^r \partial_r N 
- {1\over 2} N^{-2(\alpha-\beta) -2} s( \partial_k \partial^kz_m \partial_n N + \partial_k \partial^k z_n \partial_m N)
\eqno(2.17)$$
\par
The last term in equation (2.15) does not contribute to the terms we are computing and we find the action becomes 
$$
 {1\over 2\kappa _D} \int d^D x N^{2\beta-2\alpha-4} ( \partial_n N \partial_m N z_n  \partial_k \partial^k z_m  
 $$
 $$
 [-2(\alpha +\delta)(2\alpha +\delta)+2\alpha^2 (p+1) +2\alpha\delta +2\beta^2 (D-p-1) -4\beta +1] 
 $$
 $$
 +  z_m  \partial_k \partial^k z_m  \partial_n N \partial_n N )
 = f \int d^{p+1} x z^q\partial_k \partial^k z_q
 \eqno(2.18)$$
 where 
 $$
 f={1\over 2\kappa _D}  \int d^d y {\partial_m N \partial_m N \over d} [ 2\alpha^2 (p-1) +2\beta ^2 d -4\beta +d+1]
 = { ( \beta (d-2) +d)\over 2\kappa _D}  \int d^d y {\partial_m N \partial_m N \over d}
\eqno(2.19)$$
In carrying out these steps we  have taken $\delta=0$ and used the fact that 
$$
\int d^d y \partial_n N \partial_m N={ \delta _{mn}\over d}  \int d^d y \partial_q N \partial_q N
 \eqno(2.20)$$
\par
Combining equations (2.18)  and  (2.9) we find that the p brane moves according to the action 
$$
\int d^{p+1} x \{-f \partial_k  z^q \partial^k z_q
-{e\over 2 (q+1)!} G_ {i_1 \ldots i_{q+1}} G_ {j_1 \ldots j_{q+1}}\eta^{i_1j_1}\ldots \eta^{i_{q+1} j_{q+1} } )
\eqno(2.21)$$
which we recognise as the  the lowest order contribution to  the expected motion of the brane. It would be of interest to derive the full equations from the  action of equation (2.1). 
\par
It is interesting to calculate the charges associated with the branes. The Noether charge associated with the $n$ form gauge field in action (2.1) has the well known form 
$$
Q_q= \int d^dx \partial_\nu (F^{0\tau_1\ldots \tau_{n-1}} \Xi _{\tau_1\ldots \tau_{n-1}} \sqrt {-\det g})
\eqno(2.22)$$
where $\Xi _{\tau_1\ldots \tau_{n-1}}$ are the gauge parameters for  the gauge field. For the configurations we are interested in we take  the  $\Xi _{\tau_1\ldots \tau_{n-1}}$ to only have world volume components and then the remaining index on the field strength is $r$. As a result the charge is equal to 
$$
Q_q= - \int d^{p+1} x \int d^d y \partial_r(F_{0 r i_1\ldots i_{n-1}}  \Xi _{i_1\ldots i_{n-1}} N^{-4\beta -2(n-1)\alpha})
$$
$$
=- V_p \int_{S^{d-1}} d\Omega^{d-1} r^{d-1} F_{0 r i_1\ldots i_{n-1}}  \Xi _{i_1\ldots i_{n-1}} N^{-4\beta -2(n-1)\alpha})
\eqno(2.23)$$
where we have used the from of the metric of equation (2.2),   $V_p $  is the volume of the $ p$ brane and $S^{d-1}$ is the sphere at infinity in the directions transverse to the brane. 
\par
For the brane supporting $p$ form we find, using equation (2.3),  we find the well known charge 
$$
Q_p=  - (-1)^p V_p \int d\Omega^{d-1} r^{d-1}\partial_r N^{-1} \epsilon_{i_1\ldots i_p}  \Xi _{i_1\ldots i_{p}} N^{-4\beta -2p\alpha}
\eqno(2.24)$$
However, for the $q$ for gauge field we find, using equation (2.7), that the charge is given by 
$$
Q_q=  -  V_p \int d\Omega^{d-1} r^{d-1}\partial_r N^{-1} G_{0i_1\ldots i_p}  \Xi _{i_1\ldots i_{p}} N^{-4\beta -2q\alpha}
\eqno(2.25)$$
In both of these expressions we find the factor 
$$
\int d\Omega^{d-1} r^{d-1}\partial_r N^{-1}  N^{-4\beta -2p\alpha}
\eqno(2.26)$$
which if we take the expression for $N$ above equation (2.3) we find is equal to $-Q$ times the volume of the $d-1$ dimensional sphere. 
This illustrates the fact that we are dealing with a large gauge transformation with non-vanishing effects at infinity. In an interesting paper [45] the supergravity solutions for the M5 and D3 branes which have non-zero world volume fields was found. Using these results one could explicitly calculate the charges corresponding to the world volume fields. 
\par
Thus the moduli contained in the gauge transformation of equation (2.5) lead to a charge associated with the $q+1$ form gauge field in addition to the charge of the $p$ brane. For example the D branes can carry a charge associated with the world volume  vector fields on the brane in addition to the charge supporting the brane. This is the brane analogue of the magnetic monopole acquiring an electric charge. 
\par
One can also apply similar techniques to Taub-Nut solution in gravity. One could  find its moduli and use these to find its motion. This has indeed been carried out at the linearised level in reference [46] and it would be interesting to extend this analysis further. 
\par
The motion of the brane can not be specified by  the $z^n$ coordinates, the usual coordinates of spacetime  alone,   it requires the additional coordinates $B_{i\ldots}$ coming from the other moduli.  Taking the spacetime to be the coordinates required to uniquely specify events and so describe the motion we should regard the brane as moving in a spacetime with coordinates $z^n, B_{i\ldots }$ which is  an extension of the spacetime we are normally use. Thus we follow a similar path to that for the monopole. It would be interesting to derive brane scattering using this approach and quantise the brane from this perspective. In this last step   we would have a  wavefunction that depends on all of these coordinates, that is, $\Psi (z^n, B_{i\ldots })$. 
 \par
 An important  difference with the monopole is that for branes all of the moduli arise as large gauge transformations,  including  the coordinates of spacetime. Given one can have branes with different values of $p$ and with different transverse directions one finds that all of our usual spacetime  arises  as moduli.  We can also consider the moduli that arise in brane solutions that appear  in the fermionic sector of the maximal supergravity theories. This was also done in reference [6] for the breaking of local supersymmetry. One finds moduli that are spinors which are needed to describe the motion of the branes. We can interpret these as additional Grassmann odd  coordinates in an enlarged spacetime and they are in fact just the  Grassmann odd coordinates of the Salam-Strathdee superspace. The difference with our previous considerations is that, as is well known, these coordinates can be eliminated to have a theory  on a spacetime without them.

\medskip 
{\bf 3. Spacetime in E theory}
\medskip
E theory is the non-linear realisation of the semi-direct product of the Kac-Moody algebra $E_{11}$ with its vector representation $l_1$, denoted 
$E_{11}\otimes_s l_1$ [20,21]. It contains an infinite number of fields which are contained  in the $E_{11}$ part. These depend on an infinite number of coordinates contained in the $l_1$ part. Indeed the non-linear realisation is constructed from the group element $g=e^{z^A l_A}e^{A_{\alpha }R^{\alpha}}$ where $R^{\alpha}$ and   $l_A$ are the generators of $E_{11}$ and $l_1$ respectively.  The maximal supergravity theories emerge if we take the fields to depend only on the usual coordinates which appear at the very lowest level. The full non-linear equations of motion that follow from the non-linear realisation have been computed in eleven dimensions at low levels in eleven [22,23] five dimensions [22] and seven dimensions [24]. In the eleven dimensional case this has been extended to including  the  dual graviton [25,26]. There is little doubt that  one finds from the non-linear realisation not only the maximal supergravity theories in all its known dimensions but also all the gauged maximal supergravity theories [27]. Since all the low energy effects of the superstring  theories are contained in the maximal supergravity theories  we can think of E theory as the low energy effective action of string theory. For a review see [28]
\par
The field content in E theory is relatively well understood, it contains the fields of the supergravity theories as well as their duals, of which there are an infinite number [29]. It also contains  additional fields required for the gauged supergravity theories as well as further fields whose role  is not yet known. The irreducible representation corresponding to the non-linear realisation has been computed and  the physical degrees of freedom are  only those of the supergravity theories and so while there are an infinite number of fields there are only the  degrees of freedom  of the maximal supergravity theories [30,31]. 
\par
Clearly the coordinates in E theory have to be linked to the $E_{11}$ symmetry and they were  chosen to belong to  the vector representation, $l_1$  which is just short hand for the  first fundamental representation of  $E_{11}$ [21]. This was because  its lowest components transform under GL(11) in just such a way as to be identified with the usual coordinates of spacetime. Considering an  infinite number of coordinates is intimidating and what is the meaning of the higher level coordinates has so far been unclear. Also taking the fields to just depend on the lowest level coordinates so as  to recover the supergravity theories seemed an arbitrary step that did not follow from the theory. 
\par
The maximal supergravity theories, and so  E theory,   have many  solutions. In eleven dimensions we have the two brane and five brane, the pp wave,  the K-K monopole solutions as well as others. The  brane charges associated with these solutions are easily seen to be contained at low levels in the vector representation of $E_{11}$  and it is reasonable to expect that the higher level members of  this representation will correspond to   the charges of  additional solutions. As such we can think of the vector representation as containing all the brane charges.  Due to the way they are encoded in the non-linear realisation the coordinates are also in a one to one correspondence with these generators and so the brane charges and the coordinates are also in such a one to one correspondence.  There is also a map from the vector representation to the  fields, although this is more subtle. 
\par
 The existence of brane charges beyond those that appear as central charges  in the supersymmetry algebra was suggested  from dimensional reduction considerations and U duality [32]. The vector representation contains an infinite number of charges and so we should have an infinite number of brane charges most of which will have index structures that more complicated than those carried by forms [33]. Indeed there is a partial formulae for what the half BPS solutions are and also a formula for the tensions of these branes [34]. These have become known as exotic branes. Thus we can expect that there really are an infinite number of brane solutions with associated charges. 
 \par
The dynamics of the fields in E theory follow from the fact that it is a non-linear realisation and as we mentioned above this has been carried out at low levels. The fields that occur in the equations of motion  have gauge symmetries under which the equations of motion are invariant. Indeed  by  reversing what is  the usual procedure one can deduce the gauge symmetries and this has been done at low levels [22,23], see [35] for a discussion of this point. It has been proposed that the gauge symmetries have a parameter that is in the vector representation [36]. As such  they are in the same representation as the coordinates and, just like in general relativity, we can expect the coordinates to shift under these local parameters and so be unphysical. 
 \par 
 The above connections are  illustrated in the table below. In the first row we find the generators in the vector representation at low levels. We can take these to be the brane charges. In the second and third rows we show the coordinates and the gauge transformations respectively.  The table  illustrates  the fact that the brane charges, the coordinates and  the gauge parameters are all in a one to one correspondence.  
  $$
 \matrix {{\rm brane \ charges }& P_a & Z^{a_1a_2}     &Z^{a_1\ldots a_5}&Z^{a_1\ldots a_8}, \ Z^{a_1\ldots a_7,b}&\ldots\cr  
 {\rm coordinates} & x^a & x_{a_1a_2}     & x_{a_1\ldots a_5}& x_{a_1\ldots a_8}, \ x_{a_1\ldots a_7,b}&\ldots\cr 
   {\rm  gauge \ transformations  }&\xi^a & \Lambda_{a_1a_2} & \Lambda^{a_1\ldots a_5}&\Lambda^{a_1\ldots a_8}, \ \Lambda^{a_1\ldots a_7,b}&\ldots\cr 
 {\rm field} & h_a{}^b & A_{a_1a_2a_3} & A_{a_1\ldots a_6} & h_{a_1\ldots a_8, b} & \ldots  \cr }
  \eqno(3.1)$$
In the final row we find the fields which lead to the above quantities if we delete an index. 
 \par
As we have explained in the previous section  the well known brane solutions of the maximal supergravity theories have moduli. By taking these to depend on the world volume coordinates of the brane we have recovered at lowest order the dynamics of the branes in an enlarged  spacetime. The  usual  coordinates  of spacetime arise from the spontaneous breaking of local translations and the associated moduli appear in their parameters. There are additional coordinates that arise from the spontaneous breaking of gauge symmetries and the moduli contained in their parameters.  The branes can arise in different configurations and so one has such usual and additional coordinates in all directions and not just those in the directions picked out by a given brane.  From the graviton we find the spacetime coordinates $x^\mu$ while for  the two brane and five brane in M theory one would require the coordinates $x_{a_1a_2}$ and $x_{a_1\ldots a_5}$. If one continued  this process for all the infinite number of solutions to the maximal supergravity theories  one would arrive at an infinite dimensional spacetime which, at least at low levels,  agrees with the spacetime incorporated in E theory.  Thus even without thinking of formulating the maximal supergravity theories from an E theory perspective one would have to introduce the spacetime proposed in E theory. 
\par
What makes this different to the case of the monopole is that all the moduli that arise in supergravity theories are contained in the local transformations of the fields and this includes the moduli associated with the graviton and so local translations. As such all the coordinates which describe the motion of the branes arise in a uniform way.   In E theory all the fields  are connected by $E_{11}$ transformations, they transform  the graviton into  the three form, the three form into the six form gauge fields  etc and visa-versa. They are transform all  brane charges and so the gauge transformations and their moduli into each other, indeed they all belong to the vector representation. From the perspective of the low energy dynamics the moduli  become the  coordinates of spacetime and as such it is consoling to see that in the non-linear realisation of E theory the spacetime coordinates and the moduli both  belong to the  vector  representation. 
 \par
If one just wants to account for the behaviour of the degrees of freedom that are  obviously in the supergravity solutions, that is, not consider the solitons then there is no need to introduce the additional coordinates and one can just use the usual coordinates of spacetime. Seen from this view point when we recover the dynamics of the supergravity fields in E theory by keeping only the usual coordinates of spacetime it  is the correct thing to do. This breaks the $E_{11}$ symmetry but this is to be expected as one is deliberately ignoring the solitons  that are associated with the additional coordinates. 
 \par
 We are dealing with symmetries that are spontaneously broken and so we would expect massless excitations. in view of  the Goldstone  theorem. However, the original Goldstone theorem of Goldstone, Salam and Weinberg applies to spin zero scalars which spontaneously break  a rigid symmetry. It is almost always the case that the dynamics of the massless degrees of freedom can be described by a non-linear realisation, see section 13.2 of the book [19] for a review. In the situation we have here we have higher spin fields that transform under a local symmetry.  Nonetheless branes spontaneously break translation,  Lorentz symmetry and other symmetries and their dynamics can be derived from  a non-linear realisation.  For the case of a bosonic brane the $ISO(1,D-1)$ symmetry is broken to  $SO(p+1)\otimes SO(D-p-1)$ and     their well known dynamics has been derived by taking the non-linear realisation of the former symmetry with the latter subgroup  [37].  
 \par
The branes of the maximal supergravity theories can be constructed using a non-linear realisation of $I_c(E_{11})\otimes l_1$ where $I_c(E_{11})$ is the Cartan involution subalgebra of $E_{11}$ [38,39,40]. We note that the algebra $I_c(E_{11})$ was the local subalgebra used in the non-linear realisation from which the maximal supergravity theories arise. What local subalgebra one takes depends on the brane being considered. 
In this construction the coordinates enter in the same way as they do in the non-linear realisation that leads to the maximal supergravity theories, namely in the group element which contains the part $e^{x^Al_A}$.  The brane world volume fields appear  as  one of the coordinates in this extended spacetime. As such the usual coordinates of spacetime and the brane world volume fields  arise in the same  way as they do in moduli approach we discuss in the previous section. It would be good to have a generalised Goldstone theorem that covers these type of cases. 
\par
There is another way to view the introduction of the additional coordinates, namely we can construct the irreducible representations of $I_c(E_{11})\otimes l_1$ [30] mimicking the Wigner procedure for the Poincare group $SO(1,D-1)\otimes_s T^D$. Rather than consider just the momenta as Wigner did one takes the collection of branes charges in the vector representation. Among these one selects the charge, or charges,  of interest makes  a rest frame choice of interest, finds the little algebra  and takes a representation of this. These states are  then  boosted by the generators of  $I_c(E_{11})$ not in the little algebra. In this step the dependence on the additional coordinates inevitable appear, see equation (6.7) and the discussion above it of reference [30]. For the massless point particle the little algebra is $I_c(E_9)$ and the states contain the degrees of freedom of the maximal supergravity theories. 
\par
An infinite set of  covariant quantities was constructed from the generators of the vector representation [41].  Since the representation is irreducible and these are covariant constraints we should impose them on the states that carry the representation [41]. If one takes the decomposition of $I(E_{11})$ from which the IIA theory emerges then the lowest level such  conditions is the section condition introduced long ago in Seigel theory [42] which became known as Double Field Theory [43].  In the construction of  the maximal supergravity  theories from the  $E_{11}\otimes l_1$ non-linear realisation the section conditions are not needed and one can wonder if one should impose them on the fields that occur in the  interacting theory. The section conditions render the dependence on the additional coordinates to be rather trivial,  but as we have seen in this paper the additional coordinates  are associated with physical effects and so one should not implement  conditions that exclude these effects.


\medskip
{\bf {4. Conclusions }}
\medskip
A model of  spacetime was encoded in Newton's laws but this was replaced by the one in Maxwell's equations and it is this is the formulation that appears in  relativistic quantum field theory. Although the laws of quantum theory,  and in particular the Heisenberg uncertainty principle,  mean that it is not really possible to measure the spacetime coordinates of a particles. In Einstein's theory we find a different idea of what is spacetime, namely  it was described by  coordinates that are subject to local translations and as a result this theory   was in the doldrums up until the mid fifties. The problem was that as the coordinates were subject to local translations they   could  have no physical meaning. The question then arose as to  what quantity could one calculate that had a physical meaning and so  could be checked by a measurement.  The impasse was resolved when Felix Pirani who explained that  the way two particles moved apart as they travelled in a gravitational field was given by the Riemann tensor and as such  this could be measured [44]. The coordinates that appear in Einstein's theory and relativistic quantum field theory are really just parameters which can not be measured but with which one can compute quantities that can. Clearly the coordinates of spacetime have to be replaced in an underlying theory. The problem is what to replace them with. 
\par
Essentially all these ideas are based on the idea of particles and more precisely the particles corresponding to the fields that appear in these theories. However, field theories have solutions which can in some cases be interpreted as solitons which also have to obey the laws of quantum theory. The prototype example is the monopole that appears in spontaneously broken SU(2) Yang-Mills theory. As has been known for quite some time the motion of monopoles  can not be described in terms of just the usual coordinates of spacetime but require additional coordinates. These  arise as moduli that appear in large gauge transformations and can be used to determine the dynamics of the monopoles.  Gravity theories,  and in particular the maximal supergravity theories,  have not only particles but brane solutions and their motion can also be determined by the moduli that appear in the large local  transformations. In this way one also finds an enlargement of spacetime  is required  in order to account for the motion of the branes. 
\par
This is a very general phenomenon, fields theories  have  solutions that can be interpreted as point particles or branes and these solutions come with moduli which,  when made to depend on their world volume coordinates, can be used to give an account of the dynamics of these particles and branes in this enlarged spacetime. Thus we propose that we are really living in a world that has spacetime that is larger than we usually take it to be. 
\par
For the gravity, and in particular the maximal supergravity theories,  all the moduli appear in the parameters of    local symmetries that are spontaneously broken.  These  including  local translations  and gauge transformations.  The usual coordinates of spacetime arise from the local translations which are spontaneously broken by the position of the particle or brane while the spontaneously broken gauge symmetries lead to the additional coordinates. Thus the usual coordinates of spacetime and the additional coordinates arise in essentially the same way. 
\par
The theories we have been discussing contain the coordinates of spacetime but these are just parameters which have little physical meaning. The coordinates of the actual spacetime are related to things that can be measured,  such a the motion of particles as realised by Pirani in the context of general relativity. As we have explained the coordinates  for the motion  of particles and branes arise from spontaneously broken local symmetries. From this perspective spacetime is  not something that appears in the theory from the beginning but is something that is linked to measurable quantities. 
It would seem reasonable to think that in an underlying more fundamental theory spacetime emerges from spontaneously broken local symmetries. 
\par
Starting from the maximal supergravity theories,  one would, as we have explained in this paper, find an enlarged  spacetime once one takes account of  the brane solutions. Indeed one would find an infinite dimensional spacetime which has coordinates corresponding to all the brane charges that can arise. This is the case even if one did not take an E theory perspective, however, the spacetime one arrives at in this way is at low levels and almost certainly at higher levels the spacetime that appears in the  non-linear realisation of $E_{11}\otimes_s l_1$. Of course if one does not want to take account of the brane solutions one does not need the additional coordinates and one can take the fields  not to  depend on them. Indeed this is what one does if one wants to recover the maximal supergravity theories as they were discovered and so  this puzzling step seems from this perspective  to be the correct thing to do. 
\par
It has been proposed  by Montenon and Olive [13],  that the world had a duality symmetry that relates the particles corresponding to the fields that occur in the field theory, the so called elementary particles, to the particles that occur as solitons. The same applies to branes. From this perspective the usual coordinates of spacetime and the additional coordinates which arise from the solitons should be treated on an equal footing. Substantial parts of the  $E_{11}$ symmetry can be thought of as  duality transformations  and these do transform the usual coordinates of spacetime into the coordinates arising from the solitons.  
\par
Almost all that we see is described by two theories, the standard model and Einstein's theory of gravity. The former has the local translations as a symmetry while the latter has gauge transformations. These act in different ways, the former act on the coordinates of spacetime while the latter act on the different particles and leave spacetime alone. If one wants a unified theory the symmetries should act in the same way. In the enlarged 
spacetime  we have been discussing  the local translations act on the usual coordinates of spacetime but we also  have additional coordinates on which the gauge symmetries act. Thus the two types of symmetry act in a similar way. This is the case in E theory in that it unifies internal and spacetime symmetries both of which act on the spacetime it contains. We should think of the spacetime in E theory as an effective spacetime similar to the way we think of effective field theories. The underlying theory should have local symmetries which are spontaneously broken and in this step spacetime appears. It would be interesting to see what part the enlarged spacetime can play in cosmology. 
\eject
\medskip
{\bf {Acknowledgements}}
\medskip
I wish to thank Kevin Nguyen and Dionysis  Anninos for discussions. I also wish to thank the SFTC for support from Consolidated grants numbers ST/J002798/1 and ST/P000258/1. 

\medskip
{\bf {References}}
\medskip

\item{[1]} Ôt Hooft G. {\it Magnetic Monopoles in Unified Gauge Theories},  Nucl. Phys. B.  79, (19740),  276Ð284.
\item{[2]} A. Polyakov, {\it Particle Spectrum in the Quantum Field Theory}, PisÕma JETP. 20 (1974)  430Ð433.
\item{[3]} E. BogomolÕnyi, {\it The stability of classical solutions}, Sov. J. Nucl. Phys. 24  (1976)  449-454.
\item{[4]} M. Prasad and C. Sommerfield, {\it  An Exact Classical Solution for the Õt Hooft Monopole
and the Julia-Zee Dyon},  Phys. Rev. Lett. 35 (1975)  760-762.
\item{[5]} N. Manton, {\it A Remark on the Scattering of BPS Monopoles}, Phys.Lett. 110B (1982) 54-56; The Force Between Õt Hooft-Polyakov Monopoles // Nucl. Phys. B. 126 (1977) 525.
\item{[6]} T. Adawi, M.  Cederwall, U. Gran, Bengt E.W. Nilsson and B.  Razazneja. {\it Goldstone tensor modes}, JHEP 9902 (1999) 001, hep-th/981114. 
\item{[7]} P. Goddard and D. Olive, {\it Magnetic monopoles in gauge field theories},  Rep. Prog. Phys.,  41, (1978) 1358. 
\item{[8]} E. Corrigan and D. Olive, Nucl. Phys. B110  (1976) 237. 
\item{[9]} J. Harvey, {\it Magnetic Monopoles, Duality, and Supersymmetry}, hep-th/9603086
\item{[10]} G. Gibbons and N. Manton,  {\it  Classical and Quantum Dynamics of BPS Monopoles} Nucl. Phys. B.  274 (1986)  183.
\item{[11]} M. Atiyah and N. Hitchin, {\it The geometry and dynamics of magnetic monopoles}, Princeton University Press, 1988. 
\item{[12]} See the collected reviews in D. Olive and P. West, {\it  Dualities and Supersymmetric Theories}, Cambridge University Press, 1999. 
\item{[13]} C. Montonen and D. Olive, {\it Magnetic monopoles as gauge particles}, Phys. Lett. 72B (1977) 117. 
\item{[14]} D. Olive,The Electric and Magnetic Charges as Extra Components of Four Momentum, Nucl.Phys. B153 (1979) 1-12
 \item{[15]} P. Zizzi, {\it An Extension of the {Kaluza-Klein} Picture for the $N=4$ Supersymmetric Yang-Mills Theory}, Phys.Lett. 137B (1984) 57-61; {\it The Five-dimensional Origin Of The Monopole N=2 Supermultiplet}, Phys.Lett. 134B (1984) 197; {\it  	
A Kaluza-klein Picture Of Electric Magnetic Duality In Supersymmetry}, Nucl.Phys. B228 (1983) 229-241. 
\item{[16]} E. Witten, {\it Solutions Of Four-Dimensional Field Theories Via M Theory}, Nucl. Phys. {\bf B500} (1997) 3, hep-th/9703166.
\item{[17]} P. Howe, N. Lambert and P. West, {\it Classical M-Fivebrane Dynamics and Quantum $N=2$ Yang-Mills}, Phys.Lett. {\bf 419} (1998)
79, hep-th/9710034.
\item{[18]} N. Seiberg and E. Witten, {\it Electric-magnetic duality,
monopole condensation, and confinement in N=2 supersymmetric Yang-Mills
theory}, Nucl. Phys. {\bf B426} (1994) 19, hep-th/9407087; {\it
Monopoles, Duality and Chiral Symmetry Breaking in N=2 Supersymmetric
QCD}, Nucl. Phys. {\bf B431} (1994) 484, hep-th/9408099.
\item{[19]}  P. West, {\it Introduction to Strings and Branes}, Cambridge University Press, 2012. 
\item{[20]} P. West, {\it $E_{11}$ and M Theory}, Class. Quant.  
Grav.  {\bf 18}, (2001) 4443, hep-th/ 0104081. 
\item{[21]} P. West, {\it $E_{11}$, SL(32) and Central Charges}, Phys. Lett. {\bf B 575} (2003) 333-342,  hep-th/0307098. 
\item {[22]} A. Tumanov and P. West, {\it E11 must be a symmetry of strings and branes}, Phys. Lett. B759 (2016) 663, arXiv:1512.01644. 
\item{[23]}  A. Tumanov and P. West, {\it E11 in 11D}, Phys.Lett. B758 (2016) 278, arXiv:1601.03974. 
\item{[24]} M. Pettit and P. West, {\it An E11 invariant gauge fixing}, Int.J.Mod.Phys. A33 (2018) no.01, 1850009, Int.J.Mod.Phys. A33 (2018) no.01, 1850009.  
\item{[25} P. West, {\it Dual gravity and E11},  arXiv:1411.0920.
\item{[26]}  K. Glennon and P. West, {\it The non-linear dual gravity equation of motion in eleven dimensions}, Phys.Lett.B 809 (2020) 135714, arXiv:2006.02383. 
\item{[27]} F. Riccioni and P. West, {\it The E11 origin of all maximal supergravities}, JHEP 0707 (2007) 063; arXiv:0705.0752;
E. Bergshoeff, I. De Baetselier and T. Nutma, {\it E(11) and the Embedding Tensor}, JHEP 0709 (2007) 047, arXiv:0705.1304.
\item{[28]}  P. West, A brief review of E theory, Proceedings of Abdus Salam's 90th  Birthday meeting, 25-28 January 2016, NTU, Singapore, Editors L. Brink, M. Duff and K. Phua, World Scientific Publishing and IJMPA, {\bf Vol 31}, No 26 (2016) 1630043,  arXiv:1609.06863.  
 \item{[29]}  F. Riccioni and P. West, {\it Dual fields and $E_{11}$},    Phys.Lett.B645 (2007) 286-292,  hep-th/0612001. 
\item{[30]} P. West, {\it  Irreducible representations of E theory},  Int.J.Mod.Phys. A34 (2019) no.24, 1950133,  arXiv:1905.07324. 
\item{[31]} K. Glennon and P. West, {\it  The massless irreducible representation in E theory and how bosons can appear as spinors },     Int.J.Mod.Phys.A 36 (2021) 16, 2150096, arXiv:2102.02152. 
\item{[32]} N. Obers,Ê B. Pioline and E.Ê Rabinovici, {\it M-theory and U-duality on $T^d$ with gauge backgrounds}, {\tt hep-th/9712084}; N. Obers and B. Pioline,~ {\it U-duality and ÊM-theory, an algebraic approach}~, {\tt hep-th/9812139}; N. Obers and B. Pioline, {\it U-duality and ÊM-theory}, {\tt arXiv:hep-th/9809039}. 
\item{[33]} P. West, {\it The IIA, IIB and eleven dimensional theories and their common $E_{11}$ origin}, Nucl. Phys. B693 (2004) 76-102, hep-th/0402140; P. Cook, {\it Exotic E11  branes as composite gravitational solutions}, Clas. Quant. Grav.26 (2009) 235023, arXiv:0908.0485.  
P. Cook, {\it Bound states of string theory and beyond}, JHEP 1203 (2012) 028, arXiv:1109.6595. 
\item{[34]} P. Cook and P. West, {Charge multiplets and masses for E(11)}, JHEP {\bf 11} (2008) 091, arXiv:0805.4451. 
\item{[35]} P. West,  {\it On the different formulations of the E11 equations of motion}, Mod.Phys.Lett. A32 (2017) no.18, 1750096,  arXiv:1704.00580.
 \item{[36]} P. West,  {\it  Generalised Space-time and Gauge Transformations}, JHEP 1408 (2014) 050, arXiv:1403.6395. 
\item{[37]}  P. West, {\it Automorphisms, Non-linear Realizations and Branes}, arXiv:hep-th/0001216. 
\item{[38]} P. West, {\it Brane dynamics, central charges and $E_{11}$}, JHEP 0503 (2005) 077, hep-th/0412336.  
\item{[39]} P. West,  {\it E11, Brane Dynamics and Duality Symmetries}, Int.J.Mod.Phys. A33 (2018) no.13, 1850080, arXiv:1801.00669. 
\item{[40]} P. West, {\it A sketch of brane dynamics in seven and eight dimension using E theory}, Int.J.Mod.Phys. A33 (2018) no.32, 1850187, 
arXiv:1807.04176.
\item{[41]} P. West, {\it Generalised BPS conditions}, Mod.Phys.Lett. A27 (2012) 1250202, arXiv:1208.3397.
\item{[42]} W. Siegel, {\it Two vielbein formalism for string inspired axionic gravity}, Phys.Rev. D47
(1993) 5453, hep-th/9302036; {\it Superspace duality in low-energy superstrings}, Phys.Rev. D48 (1993) 2826-
2837, hep-th/9305073; Manifest duality in low-energy superstrings, In *Berkeley 1993,
Proceedings, Strings Õ93* 353, hep-th/9308133.
\item{[43]} O. Hohm and S. Kwak, {\it Frame-like Geometry of Double Field Theory,} J.Phys.A44
(2011) 085404, arXiv:1011.4101.
\item{[44]}  F. Pirani, {\it  On the physical significance of the Riemann tensor} . Acta Physica Polonica 15 (1956) 389. 
\item{[45]} M.  Cederwall, U. Gran, M. Holm and B. Nilsson. {\it Finite Tensor Deformations of Supergravity Solitons}, JHEP 02 (1999) 003, hep-th/9812144. 
\item{[46]} T. Maxfield and  S. Sethi {\it DBI from Gravity},     JHEP 02 (2017) 108, arXiv:1612.00427. 

\end


  {\it  The string little algebra}, (with K. Glennon), arXiv:2202.01106.

\par
We close our considerations of the one monopole motion by giving a rough and ready derivation of the motion which we will use in other cases later on in this paper. Rather than substitute the changes of equations (1.2.2) and (1.2.8) into the action we can substitute them into the equations of motion and in particular the equation of motion (1.1.3), that is, $D_0 E_i+\epsilon^{ijk} D_j B_k +e[\phi , D_i \phi ]=0$. For the change of equation (1.1.2) involving $\chi$ we see that the last two terms do not lead to any $\dot \chi$ and so one finds no change leaving us with $D_0 E_i=\partial_0 E_i= \ddot \chi { D_i \phi \over v}=0 $. As a result we conclude that  $\ddot \chi =0$. Similarly for the change of equation (1.2.8) 
involving $a^i$ we find that $D_0 E_i=\partial_0 E_i= \ddot a^j \epsilon _{ijk} B^k=0 $ and so we conclude that $ \ddot a^j =0$.